\documentclass[fleqn,twoside]{article}
\usepackage{espcrc2}

\usepackage{graphicx}
\usepackage[figuresright]{rotating}

\newcommand{\sss}{\scriptscriptstyle}

\newcommand{\AmS}{{\protect\the\textfont2
  A\kern-.1667em\lower.5ex\hbox{M}\kern-.125emS}}

\begin{document}

\title{QCD corrections to associated $t\bar th$ production at hadron colliders}

\author{S. Dawson\address{Physics Department, 
        Brookhaven National Laboratory, Upton, NY 11973, USA}
\thanks{This work is supported in part by the U.S. Department of Energy under grant DE-AC02-76CH00016.},
 \\[-76pt]\hfill BNL-HET-02/21\\ \hfill FSU-HEP-2002-1007\\ \hfill UB-HET-02-06\\[28pt]\hspace{2.2cm}
        L. Orr\address{Department of Physics and Astronomy,
        University of Rochester, Rochester, NY 14627, USA}
\thanks{This work is supported in part by the U.S. Department of Energy under grant DE-FG02-91ER40685.},  
        L. Reina\address{Physics Department, Floriday State University,
        Tallahassee, FL 32306, USA}
\thanks{This work is supported in part by the U.S. Department of Energy under grant DE-FG02-97ER41022.}
        and
        D. Wackeroth\address{Department of Physics, 
                              SUNY at Buffalo, NY 14260, USA}}
       
\begin{abstract}
We briefly present the status of QCD corrections to the inclusive total
cross section for the production of a Higgs boson in
association with a top-quark pair within the Standard Model 
at hadron colliders.
\vspace{-1pc}
\end{abstract}

\maketitle
\section{Introduction}
The search for the Higgs boson of the Standard Model (SM) is one of
the major tasks of the next generation of high-energy collider
experiments. The direct limit on the SM Higgs boson mass, $M_h$, from
LEP2 searches
\cite{lephwg1} and the indirect limit from electroweak precision
data~\cite{lepewwg} strongly suggest the existence of a light Higgs
boson, that may well be within the reach of a high-luminosity phase of
the Fermilab Tevatron $p\bar p$ collider ($M_h
{\mathrel{\raisebox{-.3em}{$\stackrel{\displaystyle <}{\sim}$}}}180$
GeV)~\cite{Carena:2002es}. However, since the dominant SM Higgs
production channels are plagued with low event rates and large
backgrounds, the Higgs boson search requires the highest possible
luminosity, and all possible production channels should be
considered. The production of a SM Higgs boson in association with a
top-quark pair, $p \bar p \to t \bar tH$, can play a role for almost
the entire Tevatron discovery
range~\cite{Carena:2002es,Goldstein:ag}. Although the $t \bar t h$
event rate is small, the signature of such events is quite spectacular
($W^+W^-b\bar bb\bar b$).  At the CERN LHC $pp$ collider, associated
$t \bar th$ production is one of the most promising processes for
studying both the Higgs boson and the top quark. This process will
provide a direct measurement of the top-Yukawa
coupling~\cite{Belyaev:2002ua,Zeppenfeld:2002ng}.
As for any other hadronic cross section, the
next-to-leading-order (NLO) QCD corrections are expected to be
numerically important and are crucial in reducing the (arbitrary)
dependence of the cross sections on the factorization and
renormalization scales.  
In the following we briefly describe the
status of NLO QCD predictions for associated $t\bar th$ production at
the Tevatron and the LHC.

\section{QCD corrections to $t \bar t h$ production}

The inclusive total cross section for $p\,p\hskip-7pt\hbox{$^{^{(\!-\!)}}$}
\rightarrow t\bar th$ at ${\cal O}(\alpha_s^3)$ can be written as:
\begin{eqnarray}\label{eq:sig}
\vspace*{-0.1cm}
&&\sigma_{\sss NLO}(p\,p\hskip-7pt\hbox{$^{^{(\!-\!)}}$}\rightarrow t\bar t h) =\sum_{ij} \int dx_1 dx_2  \\
&&{\cal F}_i^p(x_1,\mu) {\cal F}_j^{p(\bar p)}(x_2,\mu)
{\hat \sigma}^{ij}_{\sss NLO}(x_1,x_2,\mu)\,\,\,,\nonumber
\end{eqnarray}
where $ {\cal F}_i^{p, {\overline p}}$ are the NLO parton distribution
functions for parton $i$ in a proton/antiproton, defined at a
generic factorization scale $\mu_f\!=\!\mu$, and ${\hat
\sigma}^{ij}_{\sss NLO}$ is the ${\cal O}(\alpha_s^3)$ parton-level total
cross section for incoming partons $i$ and $j$, renormalized at an
arbitrary scale $\mu_r$, that we choose $\mu_r\!=\!\mu_f\!=\!\mu$.  The
NLO parton-level total cross section ${\hat \sigma}^{ij}_{\sss NLO}$ 
consists of the ${\cal
O}(\alpha_s^2)$ Born cross section and the ${\cal O}(\alpha_s)$
corrections to the Born cross section, including the effects of mass
factorization.  It comprises 
virtual and real corrections to the parton-level $t\bar th$ production
processes, $q\bar q\rightarrow t\bar th$ and $gg \to t\bar th$, and the
tree-level $qg$ initiated processes, $qg\to t\bar thq$, which are of
the same order in perturbation theory.  The main challenges in the
calculation of the ${\cal O}(\alpha_s)$ corrections come from the
presence of pentagon diagrams in the virtual part with several massive
external and internal particles, and from the computation of the real
part in the presence of infrared (IR) singularities.

\subsection{$t \bar t h$ production at the Tevatron}

The NLO inclusive total cross section for $p\bar p \to t \bar t h$
at the Tevatron center-of-mass energy $\sqrt{s_H}\!=\!2$~TeV
has been calculated independently by two 
groups~\cite{Beenakker:2001rj,Reina:2001sf,Reina:2001bc}.  The
numerical results of both calculations have been compared and they are
found to be in very good agreement.  For $p {\overline p}$ collisions
at $\sqrt{s_{H}}\!=\!2$~TeV, more than $95\%$ of the tree-level cross
section comes from the sub-process $q \bar q\rightarrow t\bar th$ and
the $gg$ initial state is numerically irrelevant. Therefore, in
\cite{Reina:2001sf,Reina:2001bc} 
when calculating $\sigma_{\sss NLO}(p\bar p\rightarrow t\bar th)$ of
Eq.~\ref{eq:sig}, we only included the $q\bar q\rightarrow t\bar th$
channel, summed over all light quark flavors. In~\cite{Beenakker:2001rj} the
${\cal O}(\alpha_s)$ corrections to both $t\bar th$ production channels
have been considered.

In~\cite{Reina:2001sf,Reina:2001bc} we have calculated the pentagon
scalar integrals as linear combinations of scalar box integrals using
the method of ~\cite{Bern:1993em}.  The real corrections are computed
using the phase space slicing method, in both the double
\cite{Harris:2001sx} and single
\cite{Giele:1992vf,Giele:1993dj,Keller:1999tf} cutoff approach.  This
is the first application of the single cutoff phase space slicing
approach to a cross section involving more than one massive particle
in the final state. As discussed in detail in~\cite{Reina:2001bc}, the
numerical results of both methods agree within the statistical errors.
In \cite{Beenakker:2001rj} the dipole subtraction formalism has been
used to extract the IR singularities of the real part.  To find
agreement between calculations based on either of the three 
very different treatments of the real IR
singularities represents a powerful check of the NLO calculations.

\begin{figure}
\includegraphics[width=14pc]{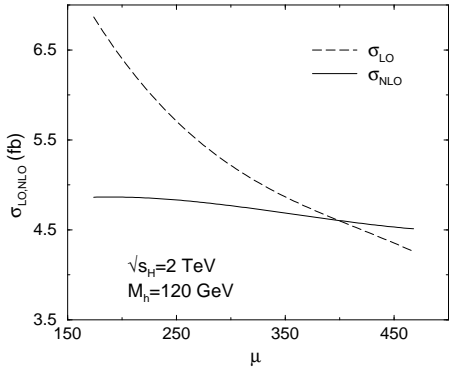}
\vspace*{-0.5cm}
\caption{\label{fg:mudep}
  Dependence of $\sigma_{\sss LO,\sss NLO}(p {\overline p}\rightarrow t
  {\overline t} h)$ on the renormalization/factorization scale $\mu$,
  at the Tevatron ($\sqrt{s_{H}}\!=\!2$~TeV), for $M_h\!=\!120$ GeV~\cite{Reina:2001sf,Reina:2001bc}.  }
\vspace*{-0.65cm}
\end{figure}

As illustrated in Fig.~\ref{fg:mudep} for $M_h=120$ GeV, 
$\sigma_{\sss NLO}(p {\overline p}\rightarrow t {\overline t} h)$ 
at $\sqrt{s}_H\!=\!2$~TeV shows a significantly
reduced scale dependence as compared to the leading-order (LO) result
and leads to increased confidence in predictions based on these
results. Over the entire range of $M_h$ accessible at the Tevatron, the
NLO corrections decrease the LO rate for 
$\mu < 2 m_t+M_h$.

\begin{figure}[htb]
{\centerline{
\setlength{\unitlength}{1cm}
\begin{picture}(7,5.4)
\put(-1.4,-14.2){\includegraphics{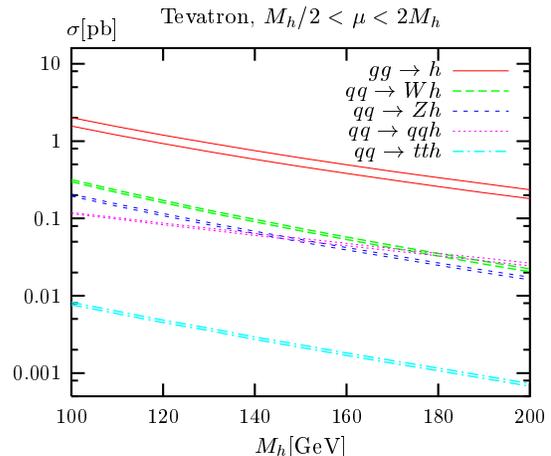}}
\end{picture} }}
\vspace*{-0.5cm}
\caption{$\sigma_{\sss NLO,\sss NNLO}$ for SM Higgs production
processes at the Tevatron ($\sqrt{s_{H}}\!=\!2$~TeV) 
as a function of $M_h$. 
For $p\bar p \to t\bar th$ the renormalization/factorization
scale is varied between $m_t+M_h/2 \!<\! \mu \!<\! 2 m_t+M_h$.}
\label{fg:tevamudep}
\vspace*{-0.65cm}
\end{figure}

To determine the potential of the Tevatron to search for the SM Higgs
boson, the predictions for all Higgs production processes have to be
under good theoretical control.  In Fig.~\ref{fg:tevamudep} we show
the scale dependence of the inclusive total cross sections at NLO
(next-to-NLO in case of $gg\to h$) for the main SM Higgs production
processes, including associated $t\bar th$ production, in the $M_h$
range accessible at the Tevatron.  The $p\bar p \to b\bar bh$ cross
section is only known at LO and thus not shown in
Fig.~\ref{fg:tevamudep}. It is comparable in size to the $t\bar th$
cross section for $\mu\!=\!m_t$ and $M_h\!=\!$120 GeV.  As illustrated in
Fig.~\ref{fg:tevamudep}, the QCD NLO calculations for the $q\bar q\to
Wh,Zh$ \cite{Han:1991ia}, $qq\to qqh$ \cite{Han:1992hr} and $q\bar q
\to t\bar t h$ \cite{Beenakker:2001rj,Reina:2001sf,Reina:2001bc}
processes provide reliable predictions for the inclusive total cross
sections at the Tevatron.  However, the NLO corrections to the $gg\to
h$ cross section~\cite{Dawson:1990zj,Djouadi:1991tk} are large (up to
$\sim 100\%$), and $\sigma_{\sss NLO}$ still strongly depends on the
factorization/renormalization scale. The QCD next-to-NLO (NNLO) $gg\to
h$ cross section became available
recently~\cite{Kilgore:2002yw,Anastasiou:2002yz}. As illustrated in
Fig.~\ref{fg:tevamudep}~\cite{thanks}, the NNLO corrections are
crucial to obtain reliable predictions.

\subsection{$t \bar t h$ production at the LHC}

While the observation of the $t\bar th$ process at the Tevatron will
be quite challenging, at the LHC the $t\bar th$ mode
will play a crucial role in the 110 GeV$\!\le \!M_h \!\le\!$130 GeV
mass region both for discovery and for precision measurements of the
Higgs boson couplings.
For determining
$\sigma_{\sss NLO}(pp\rightarrow t\bar th)$ at the LHC, the ${\cal
O}(\alpha_s)$ corrections to the sub-process $g g \rightarrow t
{\overline t} h$ are crucial, since $pp$ collisions at
$\sqrt{s_{H}}\!=\!14$ TeV are dominated by the $gg$ initial state.
Results for $\sigma_{\sss NLO}(pp\rightarrow t\bar th)$ at
$\sqrt{s_{H}}\!=\!14$ TeV have been provided in
\cite{Beenakker:2001rj} and results of an independent calculation will
be presented in~\cite{lhc}.  A detailed comparison of the numerical
results of the two groups is presently under way.

Although the calculation of the ${\cal O}(\alpha_s)$ corrections to $gg \to
t\bar th$ is technically similar to the one for the $q \bar q$
initiated process, new challenges arise, e.g., the occurrence of
spurious singularities in the calculation of the pentagon diagrams due
to a vanishing Gram determinant at the phase space boundaries.  The
resulting numerical instabilities can be controlled by applying an
extrapolation procedure from the numerically safe to the unsafe region~\cite{Beenakker:2001rj}.
However, we found that, also in the case of the $gg$ initiated
subprocess, the occurrence of these singularities can be avoided by
cancelling terms in the numerator against the propagators wherever
possible, after interfering the pentagon amplitude with the Born-matrix
element. We compared the numerical results of both treatments of these
spurious singularities and found agreement within the statistical
error. Details of the calculation will be presented elsewhere~\cite{lhc}.

As is the case at the Tevatron, the scale dependence of 
$\sigma_{\sss NLO}(p p\rightarrow t {\overline t} h)$
at $\sqrt{s}_H\!=\!14$~TeV is strongly reduced compared to the LO
result.  At the LHC, the NLO QCD corrections enhance the LO cross
section over the entire range of values for the
renormalization/factorization scale shown in Fig.~\ref{fg:mudep},
unlike at the Tevatron where they mostly decrease the Born-level cross
section.

\end{document}